\documentclass[acmtog]{acmart}
\acmPrice{15.00}

\def\BibTeX{{\rm B\kern-.05em{\sc i\kern-.025em b}\kern-.08emT\kern-.1667em\lower.7ex\hbox{E}\kern-.125emX}}

\setcopyright{acmcopyright}

\setcitestyle{square}

\usepackage{wrapfig}
\usepackage{nicefrac}
\usepackage{mathtools}
\usepackage{graphicx}
\usepackage{booktabs} 
\usepackage{combelow} 
\usepackage[ruled,vlined,linesnumbered]{algorithm2e}
\usepackage{tabularx} 
\usepackage{colortbl} 

\usepackage{manfnt} 
\usepackage{textcomp} 

\usepackage{amsmath}
\usepackage{amsfonts}
\usepackage{amssymb}
\usepackage{amsbsy}

\usepackage[mathletters]{ucs}
\usepackage[utf8x]{inputenc}

\definecolor{derekBlue}{RGB}{144,210,236}
\definecolor{derekTableBlue}{RGB}{189,235,252}
\definecolor{iglGreen}{RGB}{153,203,67}
\definecolor{coralRed}{RGB}{250,114,104}
\definecolor{gray}{RGB}{180,180,180}
\definecolor{orange}{RGB}{0,0,0}

\SetCommentSty{commentFont}
\SetNlSty{textbf}{}{.}

\newcommand{\update}[1]{{\color{orange}{#1}}}

\newcommand{\refequ}[1] {Eq.~\ref{equ:#1}}
\newcommand{\refequnum}[1] {\ref{equ:#1}}
\newcommand{\reffig}[1] {Fig.~\ref{fig:#1}}
\newcommand{\reffignum}[1] {\ref{fig:#1}}
\newcommand{\reftab}[1] {Table~\ref{tab:#1}}
\newcommand{\refsec}[1] {Sec.~\ref{sec:#1}}
\newcommand{\refapp}[1] {App.~\ref{app:#1}}
\newcommand{\refalg}[1] {Alg.~\ref{alg:#1}}

\DeclareMathOperator{\Tr}{Tr}
\DeclareMathOperator*{\argmax}{arg\,max}
\DeclareMathOperator*{\argmin}{arg\,min}
\DeclareMathOperator*{\minimize}{minimize}

\newcommand{\R}{\mathbb{R}}

\newcommand{\vecFont}[1]{\mathsf{#1}}

\def\vn{\hat{\vecFont{n}}}

\def\vt{{\vecFont{t}}}
\def\vu{{\vecFont{u}}}

\def\vx{{\vecFont{x}}}

\def\vz{{\vecFont{z}}}

\newcommand{\matFont}[1]{\mathsf{#1}}
\def\mA{{\matFont{A}}}
\def\mB{{\matFont{B}}}

\def\mF{{\matFont{F}}}

\def\mM{{\matFont{M}}}

\def\mR{{\matFont{R}}}

\def\mW{{\matFont{W}}}
\def\mX{{\matFont{X}}}
\def\mY{{\matFont{Y}}}

\def\V{{\matFont{V}}}
\def\v{{\vecFont{v}}}
\def\dV{{\matFont{D}}}
\def\dv{{\vecFont{d}}}
\def\dVs{{\matFont{D}}}
\def\U{{\widetilde{\matFont{V}}}}
\def\u{{\widetilde{\vecFont{v}}}}
\def\dU{{\widetilde{\matFont{D}}}}
\def\du{{\widetilde{\vecFont{d}}}}
\def\dUs{{\widetilde{\matFont{D}}}}
\def\SO3{{\text{SO}(3)}}
\def\l1{{\ell^1}}
\def\P{{\vecFont{p}}}
\def\dPs{{\matFont{Q}}}
\def\Q{{\widetilde{\vecFont{p}}}}
\def\dQs{{\widetilde{\matFont{Q}}}}

\def\arap{{\textsc{arap}}\xspace}
\def\cubeness{{\textsc{cubeness}}\xspace} 

\begin{document}

\title{Cubic Stylization}

\author{Hsueh-Ti Derek Liu}
\affiliation{
	\institution{University of Toronto}
	\streetaddress{40 St. George Street}
 	\city{Toronto}
 	\state{ON}
 	\postcode{M5S 2E4}
 	\country{Canada}}
\email{hsuehtil@cs.toronto.edu}
\author{Alec Jacobson}
\affiliation{
	\institution{University of Toronto}
	\streetaddress{40 St. George Street}
 	\city{Toronto}
 	\state{ON}
 	\postcode{M5S 2E4}
 	\country{Canada}}
\email{jacobson@cs.toronto.edu}

\acmJournal{TOG}
\acmYear{2019}\acmVolume{38}\acmNumber{6}\acmArticle{197}\acmMonth{11} \acmDOI{10.1145/3355089.3356495}



\begin{teaserfigure}
  \centering
  \includegraphics[width=7.0in]{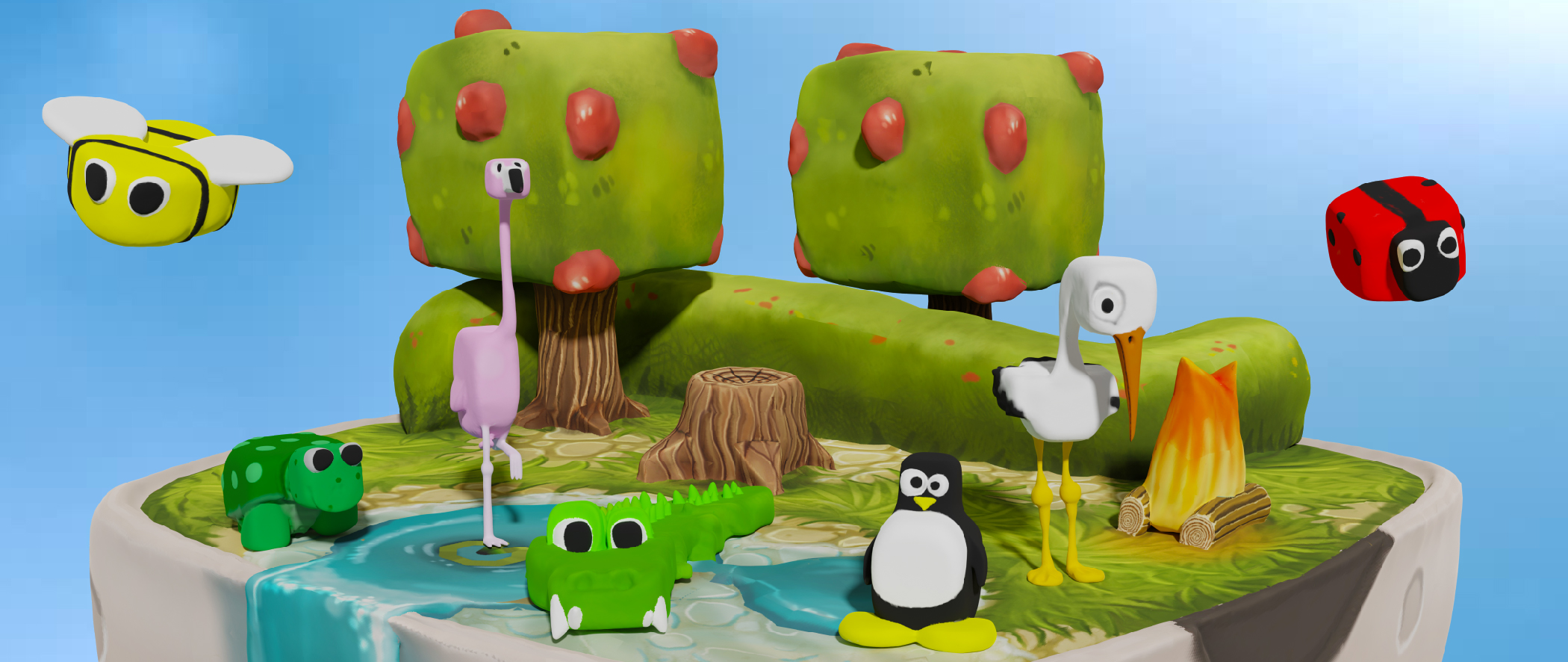}
  \caption{\emph{Cubic stylization} deforms a given 3D shape into the style of a cube while maintaining textures and geometric features. This can be used as a non-realistic modeling tool for creating stylized 3D virtual world. We obtain 3D assets from \texttt{sketchfab.com} by smeerws and Jes\'us Orgaz licensed under CC BY 4.0.}
  \label{fig:teaser}
\end{teaserfigure}

\begin{abstract}
We present a 3D stylization algorithm that can turn an input shape into the style of a cube while maintaining the \emph{content} of the original shape.
The key insight is that cubic style sculptures can be captured by the \textit{as-rigid-as-possible} energy with an $\l1$-regularization on rotated surface normals.
Minimizing this energy naturally leads to a detail-preserving, cubic geometry.
Our optimization can be solved efficiently without any mesh surgery. 
Our method serves as a non-realistic modeling tool where one can incorporate many artistic controls to create stylized geometries. 
\end{abstract}

\maketitle
\section{Introduction}
\begin{figure}[b]
    \centering
    \vspace{-5pt}
    \includegraphics[width=3.33in]{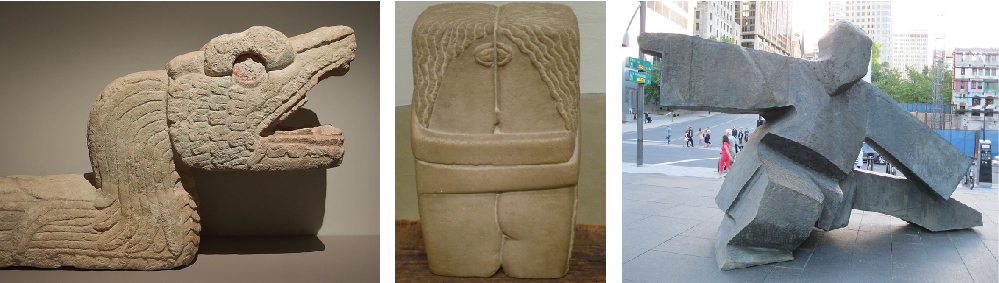}
    \caption{The cubic style have been attracting artists' attention over centuries, such as the \textit{Serpend \`a' Plumes} found in Chich\'en Itz\'a (left), \textit{The Kiss} by Constantin Br\^ancu\cb{s}i (middle), and the \textit{Taichi} by Ju Ming (right). We obtain images from \texttt{wikimedia.com} photographed by Jebulon under CC0 1.0, from \texttt{flickr.com} by Art Poskanzer under CC BY 2.0, and from \texttt{wikimedia.com} by Jeangagnon under CC BY-SA 3.0.}
    \label{fig:sculptures}
    \vspace{-5pt}
\end{figure}
The availability of image stylization filters and \textit{non-photorealistic rendering} techniques has dramatically lowered the barrier of creating artistic imagery to the point that even a non-professional user can easily create stylized images.
In stark contrast, direct stylization of 3D shapes or \emph{non-realistic modeling} has received far less attention.
In professional industries such as visual effects and video games, trained modelers are still required to meticulously create non-realistic geometric assets.
This is because investigating geometric styles is more challenging due to arbitrary topologies, curved metrics, and non-uniform discretization.
The scarcity of tools to generate artistic geometry remains a major roadblock to the development of geometric stylization.

\begin{figure}
    \centering
    \includegraphics[width=3.33in]{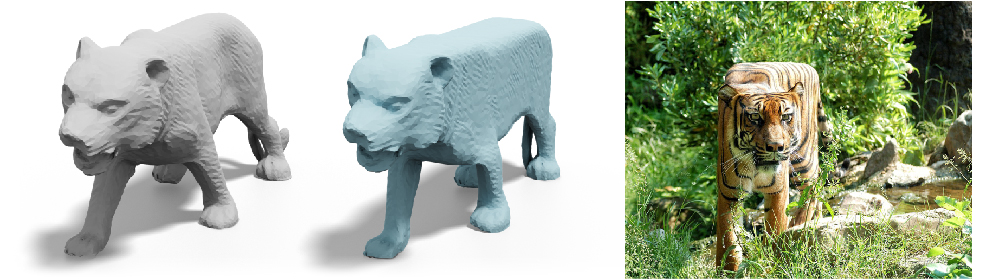}
    \caption{A digital art -- Anicube -- by Aditya Aryanto produces cubic style images (right). Our method takes an input tiger (left) and outputs a ``3D anicube'' tiger while maintaining geometric details (middle). \textcopyright Aditya Aryanto (right). Used under permission. }
    \label{fig:anicube}
    \vspace{-5pt}
\end{figure}

\begin{figure}
    \centering
    \includegraphics[width=3.33in]{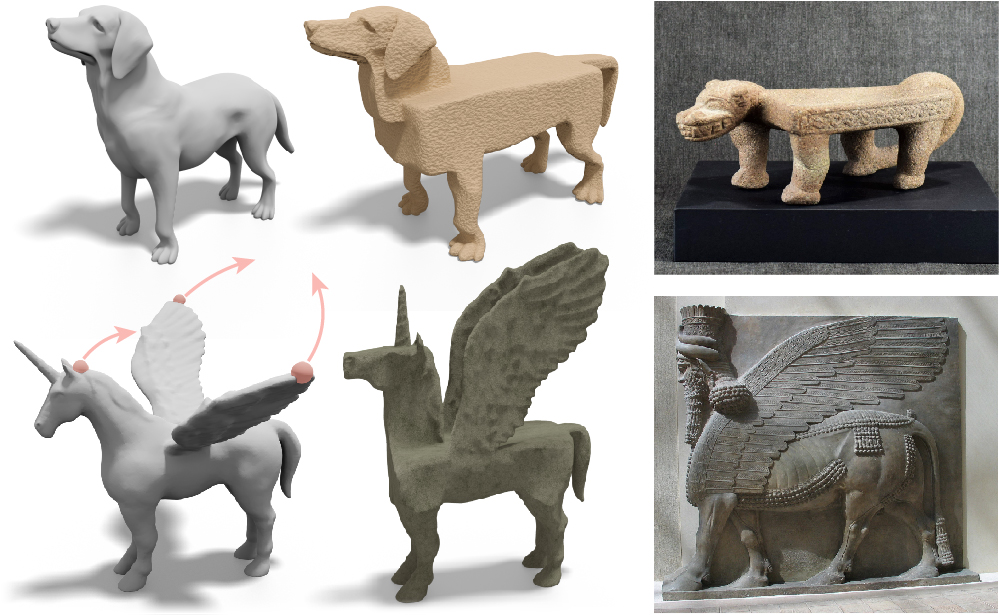}
    \caption{One can control the cubic stylization by incorporating constraints. For instance, we can fix some parts of a shape to mimic the style of a Jaguar metate from ancient Costa Rica (top) or add point constraints to mimic the Assyrian Lamassu wall sculpture (bottom). \textcopyright Antiques \& Artifacts LLC (top). Used under permission.}
    \label{fig:constraints}
    \vspace{-5pt}
\end{figure}

In this paper, we focus on the specific style of \emph{cubic} sculptures. 
The cubic style is prevalent across art history, for instance the ancient sculptures from the post-classic era (900-1250 CE), Maya sculptures, block statues in Egypt, and modern abstract sculptures such as the ones from Constantin Br\^ancu\cb{s}i and Ju Ming (\reffig{sculptures}).  
In addition, the cubic style is a popular digital art, such as the award-winning \emph{Anicube} by Aditya Aryanto (\reffig{anicube}).
Complementing their presence in art, cubic shapes also present themselves in fabrication and furniture purposes (\reffig{constraints}).
\begin{figure}
    \centering
    \includegraphics[width=3.33in]{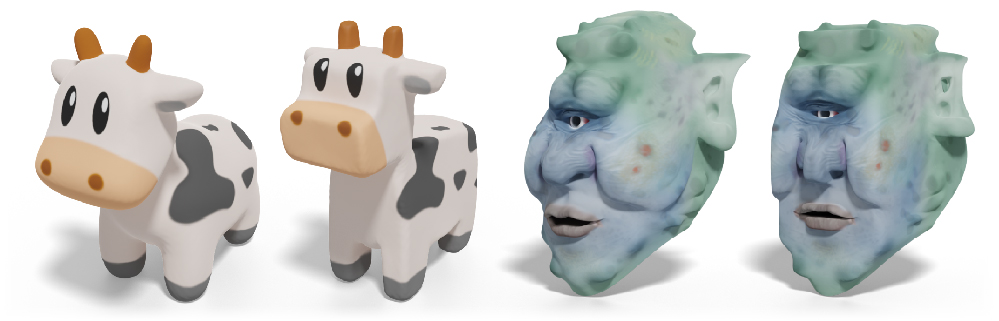}
    \caption{Our cubic stylization requires no remeshing, thus vertex attributes such as textures are preserved during the optimization. \update{Our \arap term encourges locally isometric deformations to help maintain nice textures.}}
    \label{fig:texturePres}
    \vspace{-5pt}
\end{figure}
We contribute to the rich history of cubic sculpting by providing a stylization tool that takes a 3D shape as input and outputs a deformed shape that has the same style as cubic sculptures.

We present \emph{cubic stylization} which formulates the task as an energy optimization that naturally preserves geometric details while cubifying a shape.
Our proposed energy combines an \emph{as-rigid-as-possible} (\arap) energy with an $\l1$ regularization. 
This energy can be minimized efficiently using the local-global approach with \textit{alternating direction method of multipliers} (ADMM).
This variational approach affords the flexibility of incorporating many artistic controls, such as applying constraints, non-uniform cubeness, and different global/local cube orientations (\refsec{results}).
Moreover, our method requires no remeshing (\reffig{texturePres}) and generalizes to polyhedral stylization (\reffig{polyCF}).
%
%
Our proposed tool for non-realistic modeling goes beyond the 2D stylization and opens up the possibility of, for instance, creating non-realistic 3D worlds in virtual reality (\reffig{teaser}).

\section{Related Work}\label{sec:relatedWork}
Our work shares similar motivations to a large body of work on image stylization \cite{kyprianidis2013state}, non-photorealistic rendering \cite{gooch2001non}, and motion stylization \cite{hertzmann2009realistic}. While their outputs are images or stylized animations, we take a 3D shape as input and output a stylized shape. Thus we focus our discussion on methods for processing geometry, including the study of geometric styles and deformation methods that share technical similarities. 

\paragraph{Discriminative Geometric Styles}
The growing interest in understanding geometric styles has been inspiring recent works on building \emph{discriminative} models for style analysis.
One of the main challenges is to define a similarity metric aligned with human perception.
Many works propose to compare projected feature curves \cite{li2013curve, yu2018semi}, sub-components of a shape \cite{xu2010style, lun2015elements, hu2017co}, or using learned features \cite{lim2016identifying}. 
These models enable users to synthesize style compatible scenes \cite{liu2015style} or transfer style components across shapes \cite{ma2014analogy, lun2016functionality, berkiten2017learning}. 
However, these methods are designed for discerning and transfering styles, instead of generating 3D stylized shapes directly.

\begin{figure}
    \centering
    \includegraphics[width=3.33in]{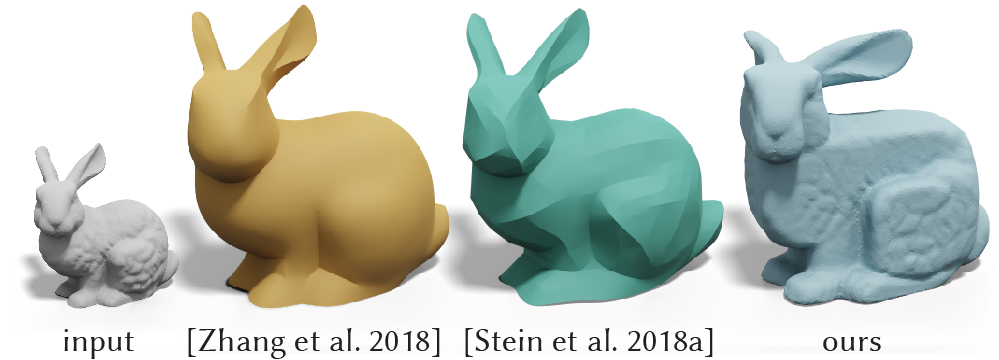}
    \caption{Our energy-based deformation shares similarities with many energy-based geometric flows and mesh filters, such as the methods of \cite{zhang2018static} and \cite{Stein2018DSF}. }
    \label{fig:comparisons}
    \vspace{-5pt}
\end{figure}

\begin{figure}
    \centering
    \includegraphics[width=3.33in]{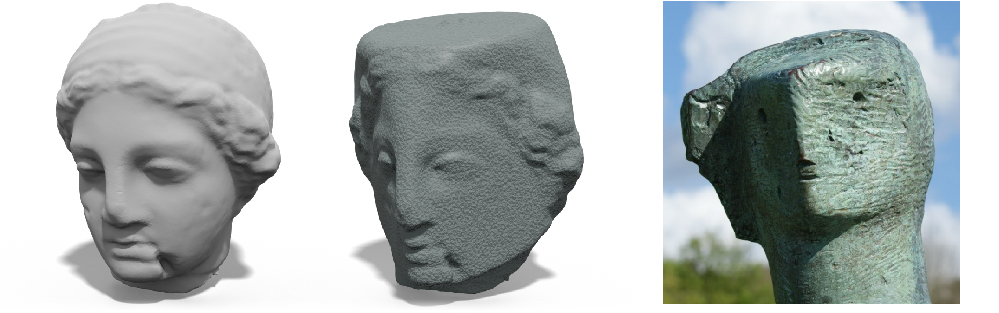}
    \caption{Cubic style sculptures are common throughout history, such as the \update{\textit{Draped Seated Woman} by Henry Moore (right)}. Our cubic stylization offers an instrument to create cubic geometry (middle). We obtain the the photo from \texttt{flickr.com} photographed by puffin11k under CC BY-SA 2.0.}
    \label{fig:HenryMoore}
    \vspace{-5pt}
\end{figure}

\paragraph{Generative Geometric Styles}
Direct 3D stylization has been an important topic in computer graphics.
Many \emph{generative} models have been proposed for producing specific styles, without relying on identifying and transferring style components from other shapes.
This includes creating the collage art \cite{gal20073d, Theobalt2007Collage}, voxel/lego art \cite{testuz2013lego, luo2015legolization}, \textit{neuronal homunculus} \cite{reinert2012homunculus}, the manga style shapes \cite{shen2012sd}, shape abstraction \cite{mehra2009abstraction, Kratt2014NRS, yumer2012co}, and \textit{bas-relief} sculptures \cite{weyrich2007digital, song2007automatic, kerber2009feature, bian2011preserving, schuller2014appearance}.
While not pitched as stylization techniques, many geometric flows and filters can also be used for creating stylized geometry, such as creating edge-preserving smoothing geometry \cite{zhang2018static}, piece-wise planar \cite{he2013mesh, Stein2018NBC} or developable shapes \cite{Stein2018DSF}, and stylized shapes prescribed by image filters \cite{liu2018paparazzi} (see \reffig{comparisons}).
Our method contributes to the field of direct 3D stylization, focusing on the style of cubic sculptures (\reffig{HenryMoore}).
\begin{figure}
    \centering
    \includegraphics[width=3.33in]{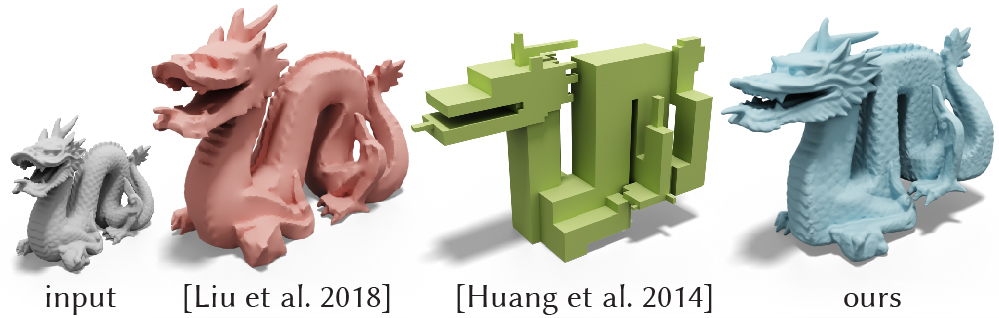}
    \caption{Paparazzi \cite{liu2018paparazzi} with image quantization and polycube method (e.g., \cite{huang2014}) can create cubic style shapes (red, green), but unlike our method (blue) they do not preserve geometric details.}
    \label{fig:polycube}
    \vspace{-5pt}
\end{figure}

\paragraph{Shape Deformation}
Many works deal with the question of how to deform shapes given modeling constraints.
One of the most popular choices is the \arap energy \cite{igarashi2005rigid, sorkine2007rigid, liu2008local, chao2010simple}, which measures local rigidity of the surface and leads to detail-preserving deformations.
Not just deformations, similar formulations to \arap can also be extended to other tasks such as constrained shape optimization \cite{bouaziz2012shape}, parameterization \cite{liu2008local}, and simulating mass-spring systems \cite{liu2013fast}.
Ever since, optimizing the \arap energy has been substantially accelerated by a large amount of work, such as \cite{kovalsky2016accelerated, rabinovich2017scalable, shtengel2017geometric, peng2018anderson, zhu2018blended}. 
%
However, having nearly interactive performance on highly detailed meshes still remains a major challenge.
An alternative strategy to speed it up is to use the hierarchical deformation which optimizes \arap on a low resolution model and then recover the original details back afterwards \cite{manson2011hierarchical}.
This class of accelerations shares similar characteristics to multiresolution modeling (see \cite{garland1999multiresolution, zorin2006modeling}).
We take advantage of the \arap energy for detail preservation and adapt the method of \citet{manson2011hierarchical} to accelerate our cubic stylization to meshes with millions of faces.

\paragraph{Axis-Alignment in Polycube Maps}
Axis-alignment is an important property for many geometry processing tasks, such as \cite{muntoni2018axis, stein2019interactive}. 
Especially, this concept is one of the main instruments in the construction of polycube maps \cite{tarini2004polycube}, including defining polycube segmentations \cite{livesu2013polycut, fu2016efficient, Zhao2018} and the cost function for polycube deformations \cite{gregson2011all, huang2014}.
Although polycube methods can obtain cubic geometry, they fail to preserve detail (\reffig{polycube}) because they are not desirable for intended applications such as parameterization and hexahedral meshing \cite{wang2007polycube, lin2008automatic, wang2008user, he2009divide, garcia2013interactive, yu2014optimizing, cherchi2016polycube, fang2016all}.

\vspace{4mm}
One tempting direction of creating cubic geometry is to use voxelization.
However, voxelization fails to capture the details depicted by the artists and cannot capture the wide spectrum of cubeness across cubic sculptures.
Another tempting direction is to recover geometric features from the polycube results.
This would lead to a multi-step algorithm and suffer from limitations of particular detail encoding schemes (e.g., bump maps). \update{Even if we stop the polycube}
\begin{wrapfigure}[10]{r}{1.05in}
	\raggedleft
	\vspace{-15pt}
	\hspace*{-0.7\columnsep}
	\includegraphics[width=1.11in, trim={3mm 0mm 1mm 0mm}]{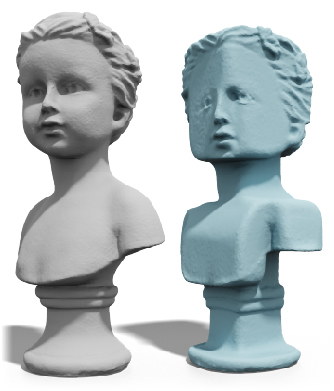} 
	\label{fig:buste}
\end{wrapfigure} 
\update{algorithm earlier such as the method of \cite{gregson2011all} to maintain details, it does not provide a satisfactory solution (see the inset for a comparison with Fig. 5 in \cite{gregson2011all}).}
More importantly, many artistic controls in \refsec{results} would be nontrivial to add on.
Building stylization on top of polycube methods would also suffer from slow performance. 
For instance, \citet{huang2014} propose a polycube method that minimizes the $\l1$-norm of the normals on the deformed \update{tetrahedral} mesh with \arap for regularization.  
Their formulation involves minimizing a complicated non-linear function and requires minutes to hours to optimize.
Thus a stylization built on top of this method would be even slower.
In contrast, our formulation is a single energy optimization which can easily incorporate many artistic controls (\refsec{results}). 
Our energy is similar to the polycube energy of \cite{huang2014} in that we also minimize the \arap energy with a $\l1$ regularization, but the key difference is that we define the $\l1$-norm on the \emph{rotated normals} of the \emph{original} mesh instead.
This allows us to optimize our energy much faster using the local-global approach with ADMM in only a few seconds (\reftab{runtime}).

\section{Method} \label{sec:method}
\begin{figure*}
    \centering
    \includegraphics[width=7.0in]{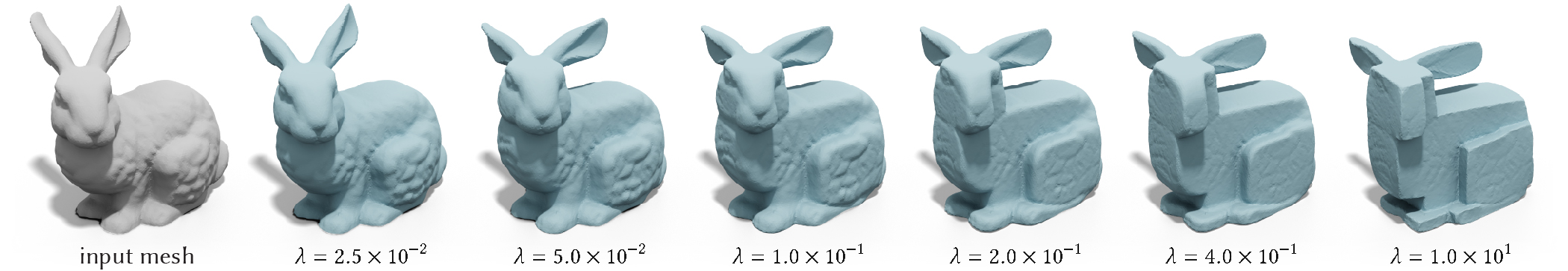}
    \caption{One can control the \cubeness by changing the $λ$ parameter \update{in \refequ{cubeEnergy}}.}
    \label{fig:changeLambda}
    \vspace{-5pt}
\end{figure*}  
The input of our method is a manifold triangle mesh with/without boundaries.
Our method outputs a \emph{cubified} shape where each subcomponent has the style of an axis-aligned cube.
Meanwhile, our stylization will maintain the geometric details of the original mesh.

Let $\V$ be a $|\V|×3$ matrix of vertex positions at the rest state and $\U$ be
a $|\V|×3$ matrix containing the deformed vertex positions. 
We denote $\dv_{ij} = [\v_j - \v_i]^{⊤}$ and $\du_{ij} = [\u_j - \u_i]^{⊤}$ be
the edge vectors between vertices $i,j$ at the rest and deformed states
respectively. 
The energy for our cubic stylization is as follows
\begin{align}\label{equ:cubeEnergy}
    \minimize_{\U, \{ \mR_i\} }\ \sum_{i∈V} \sum_{j∈ \mathcal{N}(i)} \underbrace{\frac{w_{ij}}{2}  \| \mR_i \dv_{ij} - \du_{ij} \|^2_F}_{\arap} + \underbrace{λ a_i \| \mR_i \vn_i \|_1}_{\cubeness}.
\end{align}
The first term is the \arap energy
\cite{sorkine2007rigid}, where $\mR_i$ is a $3$-by-$3$ rotation matrix,
$w_{ij}$ is the cotangent weight \cite{pinkall1993computing}, and
$\mathcal{N}(i)$ denotes the ``spokes and rims'' edges of the $i$th vertex
\cite{chao2010simple} (see the inset). 
In the second term, $\vn_i$ denotes the unit \update{area-weighted} normal vector of a vertex $i$ in
$\R^3$. The $a_i ∈ \R^+$ is the barycentric area of vertex $i$, which is crucial
for $λ$ to exhibit the similar cubeness across different mesh resolutions.
\begin{wrapfigure}[9]{r}{1.05in}
	\raggedleft
	\vspace{-6pt}
	\hspace*{-0.7\columnsep}
	\includegraphics[width=1.11in, trim={3mm 0mm 1mm 0mm}]{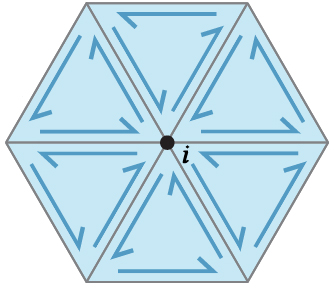} 
	\label{fig:spokesRims}
\end{wrapfigure} 
Intuitively, minimizing the $\l1$-norm of the rotated normal encourages $\mR_i
\vn_i$ to align with one of coordinate axes because $\l1$-norm encourages sparsity. 
Combining the two,
the optimal rotation $\{ \mR_i^\star \}$ would simultaneously preserve the local
structure (\arap) and encourage axis alignment (\cubeness). 

We adapt the standard local-global update strategy to optimize our energy \cite{sorkine2007rigid} (see \refalg{cubeStyle}). 
Our global step, updating $\U$, is achieved by solving a linear system, the same as the Equation 9 in \citet{sorkine2007rigid}. 
Our local step, finding the optimal rotation, is however different from the previous literature due to the $\l1$ term.

\subsection{Local-Step} \label{sec:localStep}
Our local step for each vertex $i$ can be written as 
\begin{align}\label{equ:localStep}
    \mR^⋆_i 
    & = \argmin_{\mR_i ∈ \SO3}\ \frac{ 1}{2} \| \mR_i\dVs_i \ - \dUs_i \|^2_{\mW_{i}} + λ a_i \| \mR_i \vn_i \|_1,
\end{align}
where $\mW_i$ is a $|\mathcal{N}(i)|×|\mathcal{N}(i)|$ diagonal matrix of
cotangent weights, $\dVs_i$ and $\dUs_i$ are $3×|\mathcal{N}(i)|$ matrices of
rim/spoke edge vectors at the rest and deformed states respectively. We denote $\| \mX \|^2_{\mY} = \Tr(\mX \mY \mX^⊤)$ for notational convenience.
By setting $\vz = \mR_i \vn_i$, we can \update{rewrite \refequ{localStep} as
\begin{align}\label{equ:localStep_rewrite}
    \minimize_{\vz, \mR_i ∈ \SO3}\quad &\frac{ 1}{2} \| \mR_i\dVs_i \ - \dUs_i \|^2_{\mW_{i}} + λ a_i \| \vz \|_1 \\
    \text{subject to}\quad &\vz - \mR_i \vn_i = 0. \nonumber
\end{align}
\refequ{localStep_rewrite} is a standard ADMM formulation. We solve this local step using the scaled-form ADMM updates \cite{boyd2011distributed}:}
\begin{alignat}{2}
    &\mR_i^{k+1} &&← \argmin_{\mR_i ∈ \SO3}\frac{1}{2}  \| \mR_i\dVs_i \ - \dUs_i \|^2_{\mW_{i}} + \frac{ρ^k}{2} \| \mR_i \vn_i - \vz^k + \vu^k \|^2_2 \label{equ:RStep}\\
    &\vz^{k+1} &&← \argmin_{\vz}\ λ a_i \| \vz \|_1 +  \frac{ρ^k}{2} \| \mR_i^{k+1} \vn_i - \vz + \vu^k \|^2_2 \label{equ:ZStep}\\
    &\tilde{\vu}^{k+1} &&← \vu^k + \mR_i^{k+1} \vn_i - \vz^{k+1} \label{equ:UStep}\\
    &ρ^{k+1} &&, \vu^{k+1} ← \textit{update\,($ρ^k$)} \label{equ:rhoStep}
\end{alignat}
\begin{figure}
    \centering
    \includegraphics[width=3.33in]{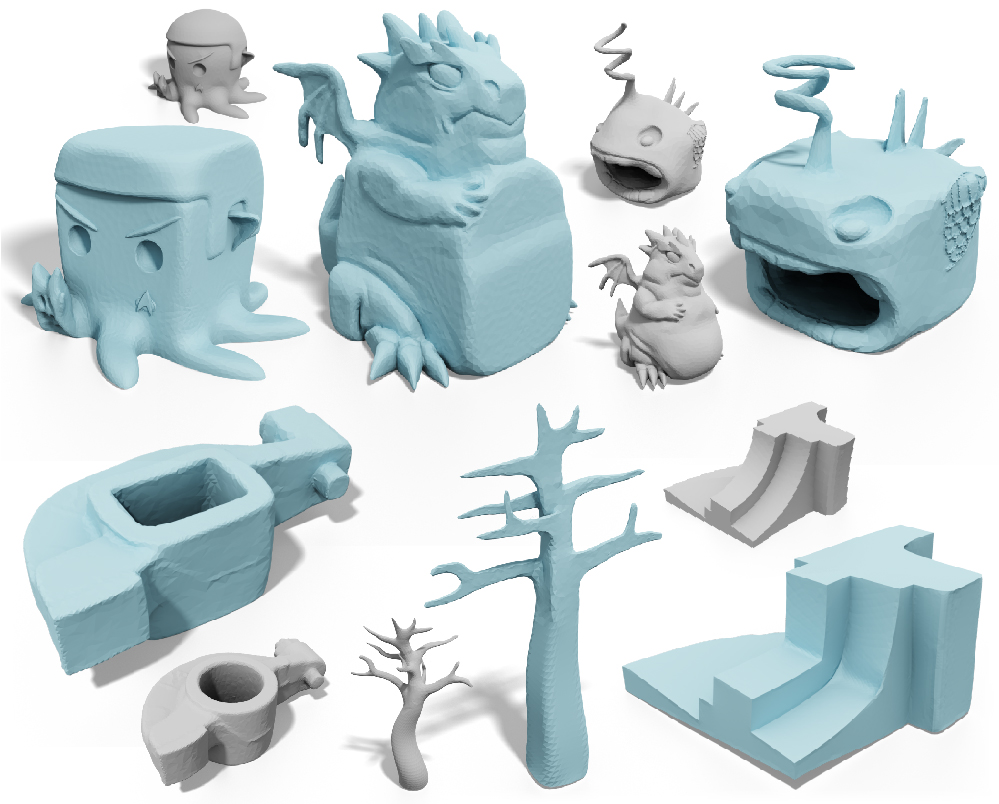}
    \caption{We turn 3D shapes into the cubic style (blue) with \refalg{cubeStyle}. \textcopyright Angelo Tartanian (top left), Splotchy Ink (top), Dan Slack (top right) under CC BY.}
    \label{fig:rawCF}
    \vspace{-5pt}
\end{figure}
\begin{figure}
    \centering
    \includegraphics[width=3.33in]{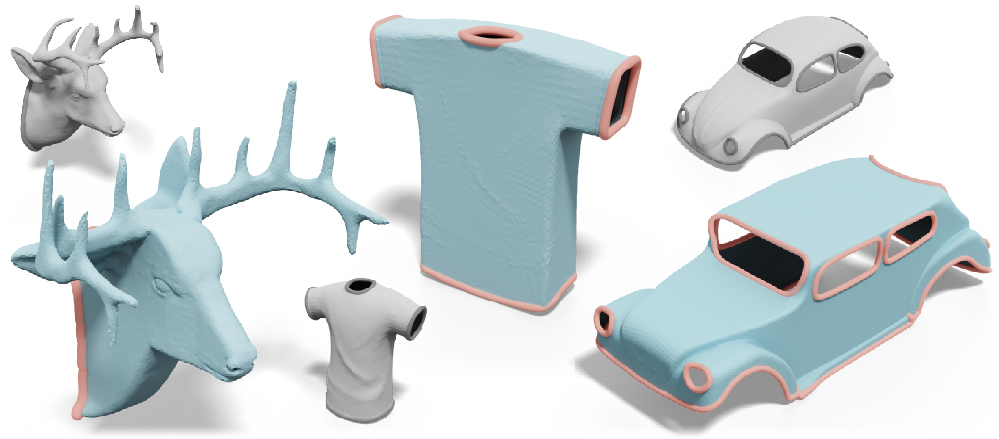}
    \caption{We can also turn meshes with boundaries (red) into the cubic style. \textcopyright Takeshi Murata (left) under CC BY.}
    \label{fig:bdMesh}
    \vspace{-5pt}
\end{figure}
where $ρ ∈ \R_+$ is the penalty and $\vu$ is the scaled dual variable.

\refequ{RStep} is an instance of the \textit{orthogonal Procrustes} \cite{gower2004procrustes}
\begin{align}
    &\mR_i^{k+1} ← \argmax_{\mR_i ∈ \SO3}\ \Tr(\mR_i \mM_i) \nonumber\\
    &\mM_i = 
    \begin{bmatrix}
        \dVs_i \quad \vn_i 
    \end{bmatrix}
    \begin{bmatrix}
        \mW_i & \\
        & ρ^k 
    \end{bmatrix}
    \begin{bmatrix}
        \dUs_i^⊤ \\ 
        (\vz^k - \vu^k)^⊤ 
    \end{bmatrix}.\nonumber
\end{align}
One can derive the optimal $\mR_i$ from the singular value decomposition of $\mM_i = \mathcal{U}_i Σ_i \mathcal{V}_i^⊤$:
\begin{align}
    \mR_i^{k+1} ← \mathcal{V}_i \mathcal{U}_i^⊤,
    \label{equ:finalRStep}
\end{align}
\update{up to changing the sign of the column of $\mathcal{U}_i$ so that $\det(\mR_i) > 0$.}

\refequ{ZStep} is an instance of the \textit{lasso} problem \cite{tibshirani1996regression, boyd2011distributed}, which can be solved with a
\textit{shrinkage} step:
\begin{align}
    &\vz^{k+1} ← \mathcal{S}_{\nicefrac{λ a_i}{ρ^k}} (\mR_i^{k+1} \vn_i + \vu^k) \label{equ:shrinkageZ}\\
    &\mathcal{S}_κ (\vx)_j= (1 − \nicefrac{κ}{|x_j|})_+\ x_j \nonumber
\end{align}
We update the penalty $ρ$ (\refequ{rhoStep}) according to Sec.~3.4.1 in \cite{boyd2011distributed}
where $\vu$ \update{needs} to be rescaled accordingly after updating $ρ$.

In short, local fitting is performed by running \refequ{finalRStep}, \refequnum{shrinkageZ}, \refequnum{UStep}, and \refequnum{rhoStep} iteratively until the norm of primal/dual residuals are small.
Warm starting the local-step parameters from the previous iteration can significantly speed up the optimization. 
Specifically, we initialize $\vz, \vu$ with zeros, and set the initial $ρ = 10^{-4}$, $ϵ^{\text{abs}} = 10^{-5}$, $ϵ^{\text{rel}} = 10^{-3}$, $μ = 10$, and $τ^{\text{incr}} = τ^{\text{decr}} = 2$ (the same notation as used \update{in Sec. 3} of \cite{boyd2011distributed}). 
Then $\vz, \vu, ρ$ are reused in consecutive iterations. Note that for extremely large $λ$ one may need to increase 
%
%
\begin{figure}
    \centering
    \includegraphics[width=3.33in]{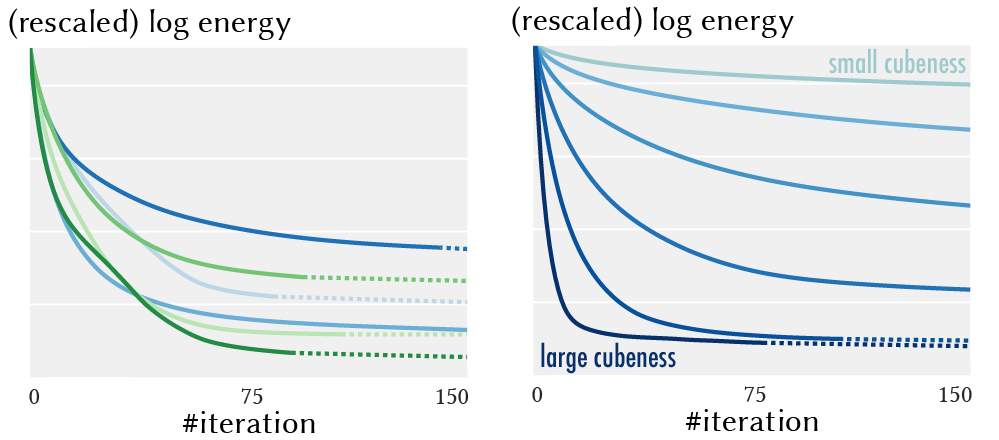}
    \caption{\update{We show the convergence behavior of different meshes in \reffig{rawCF} (left, blue), \reffig{apmCF} (left, green), and different cubenesses in \reffig{changeLambda} (right). Note that the dotted line imply the optimization has stopped.}}
    \label{fig:convergence}
    \vspace{-5pt}
\end{figure}
\begin{figure}
    \centering
    \includegraphics[width=3.33in]{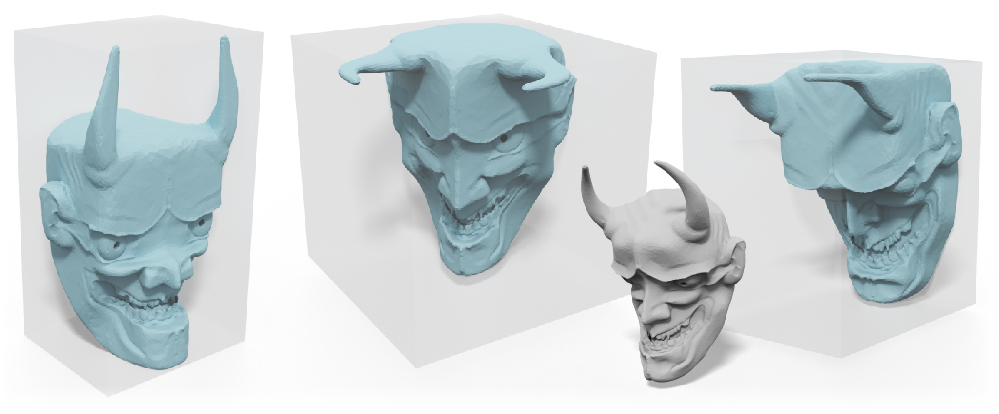}
    \caption{The global orientation of the shape influences the $\l1$ term in \refequ{cubeEnergy}. Applying different rotations to the mesh lead to different results. \textcopyright My Dog Justice under CC BY.}
    \label{fig:orientations}
    \vspace{-5pt}
\end{figure}
the initial value of $ϵ^{\text{abs}}$ accordingly in order to avoid bad local minima. We stop the 
\begin{wrapfigure}[11]{r}{0.4\linewidth}
    \vspace*{-1\intextsep}
    \hspace*{-0.5\columnsep}
    \begin{minipage}[b]{\linewidth}
    \includegraphics[width=\linewidth, trim={0mm 4mm 0mm 0mm}]{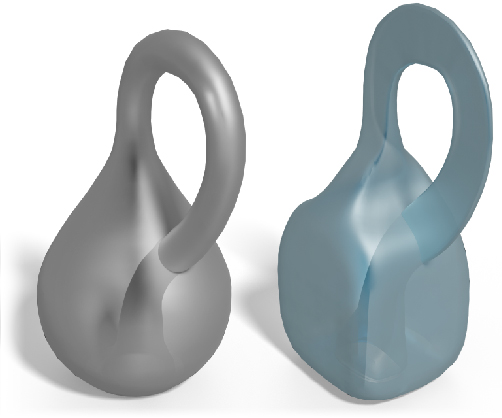}
    \caption{\update{Our method can cubify non-orientable surfaces such as the Klein bottle.}}
    \label{fig:kleinBottle}
    \end{minipage}
\end{wrapfigure}
optimization when the relative displacement, the infinity norm of relative per vertex displacements, is lower than $3×10^{-3}$ (see \reffig{convergence} for the convergence plots). More elaborate stopping criteria, such as the method of \cite{zhu2018blended}, could also be used.

%
At this point we have completed the cubic stylization algorithm summarized in \refalg{cubeStyle}, enabling us to efficiently create cubified
\update{shapes (see \reffig{rawCF}). In \reffig{bdMesh} and \reffignum{kleinBottle} we show that this formulation is applicable to meshes with boundaries and non-orientable surface respectively.}
As the \cubeness is dependent to the orientation of the mesh, one can apply different rotations to control how the stylization runs (\reffig{orientations}).
We expose the weighting $λ$ to be a design parameter controlling the cubeness of a shape (\reffig{changeLambda}). 

However, the ``vanilla'' cube stylization shares the same caveat as other distortion minimization algorithms: having slow runtime on high resolution meshes.

\IncMargin{1em}
\begin{algorithm}[t]
    \SetKwInOut{Input}{Input}
    \SetKwInOut{Output}{Output}
    \caption{$Cube\ Stylization\,(λ)$}
    \label{alg:cubeStyle}
    \Indentp{-1em}
        \Input{$\,$A triangle mesh $\V, \mF$}
        \Output{$\,$Deformed vertex positions $\U$}
        \BlankLine
    \Indentp{1em}
        $\U ← \V$\\
        \While{\textit{not converge}}{
            $\mR ← local\text{-}step\,(\V, \U, λ)$\\
            $\U ← global\text{-}step\,(\mR)$\\
        }
\end{algorithm} 
\DecMargin{1em}
\IncMargin{1em}
\begin{algorithm}[t]
    \SetKwInOut{Input}{Input}
    \SetKwInOut{Output}{Output}
    
    \caption{$Fast\ Cube\ Stylization\,(λ, m)$}
    \label{alg:fastCubeStyle}
    \Indentp{-1em}
        \Input{$\,$A triangle mesh $\V, \mF$}
        \Output{$\,$Deformed vertex positions $\U$}
        \BlankLine
        \tcp{pre-processing}
    \Indentp{1em}
        $m ← \text{target number of faces}$\\
        $\V_c, \mF_c ← \textit{edge collapses}\,(\V, \mF, m)$ \\
    \Indentp{-1em}
        \tcp{cubic stylization}
    \Indentp{1em}
        $\U_c ← \V_c$\\
        \While{\textit{not converge}}{
            $\mR\ \ ← local\text{-}step\,(\V_c, \U_c, λ)$\\
            $\U_c ← global\text{-}step\,(\mR)$\\
        }
        $\U, \mF← \textit{affine vertex splits}\,(\U_c, \mF_c)$
\end{algorithm} 
\DecMargin{1em}

\subsection{Affine Progressive Meshes}\label{sec:apm}
\citet{manson2011hierarchical} propose a hierarchical approach to accelerate \arap deformations. 
The main idea is to deform a low-resolution model and recover the details back after convergence.

Specifically, \citet{manson2011hierarchical} propose a progressive mesh \cite{hoppe1996progressive} representation which first simplifies a given mesh via a sequence of edge collapses, and then represents the mesh as its coarsest form together with a sequence of vertex splits. 
After   
\begin{wrapfigure}[11]{r}{1.37in}
	\raggedleft
    \vspace{-10pt}
	\hspace*{-0.7\columnsep}
	\includegraphics[width=1.31in, trim={6mm 0mm -1mm 0mm}]{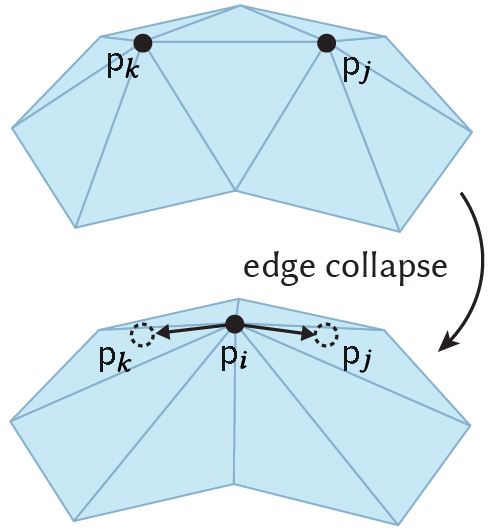} 
	\label{fig:edgeCollapse}
\end{wrapfigure} 
applying some deformations to the coarsest mesh, each ``deformed'' vertex split is computed by fitting the best local rigid transformation. This approach is suitable for deformations that 
are locally rigid (e.g., \arap), but our cubic stylization is \emph{less} rigid for larger $λ$.

So we fit the best \emph{affine} transformation in each vertex split, rather than rigid transformations. 
Specifically, in each edge collapse we store the displacement vectors from the newly inserted vertex $\P_i$ to the endpoints $\P_j, \P_k$ (see the inset) together with a matrix $\mA$:
\begin{align*}
	\mA = (\dPs_i \dPs_i^⊤)^{-1} \dPs_i.
\end{align*}
$\dPs_i$ is a $3×|\mathcal{N}(i)|$ matrix where each column is the vector from $\P_i$ to one of its one-rings neighbors $\mathcal{N}(i)$.
If $(\dPs_i \dPs_i^⊤)$ is singular (e.g., in planar regions), we remedy the issue with the Tikhonov regularization \cite{tikhonov2013numerical}.
Then $\mA$ is used to computed the deformed displacements for each vertex split as
\begin{align*}
	\Q_j - \Q_i = \dQs_i \mA^⊤ (\P_j - \P_i),
\end{align*}
where \update{$\Q_i$ denotes the position of vertex $i$ in the cubified coarsened shape, and} $\dQs_i$ is a $3×|\mathcal{N}(i)|$ matrix containing vectors from $\Q_i$ to its one-rings neighbors. 

\begin{figure}[b]
    \centering
    \vspace{-5pt}
    \includegraphics[width=3.33in]{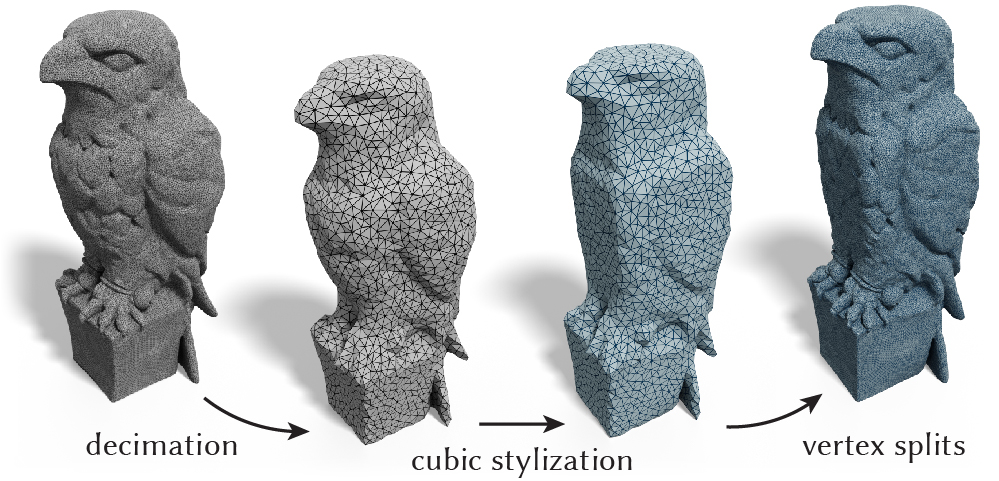}
    \caption{Affine progressive meshes allow us run cubic stylization on a low-resolution model and then recover original details when converged. \textcopyright Colin Freeman under CC BY.}
    \label{fig:apm}
    \vspace{-5pt}
\end{figure}
\begin{figure}
    \centering
    \includegraphics[width=3.33in]{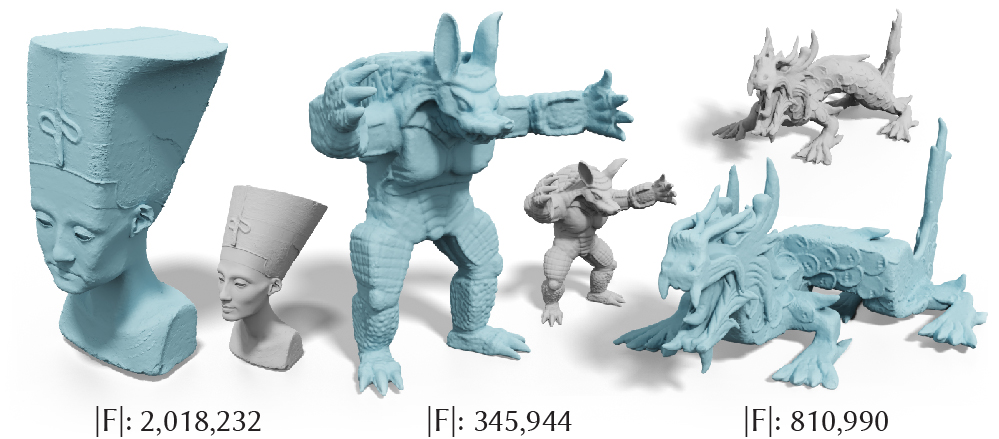}
    \caption{With the affine progressive meshes, we can scale the cubic stylization to meshes with millions of faces. The Nefertiti mesh (left) was scanned by Nora Al-Badri and Jan Nikolai Nelles from the Nefertiti bust.}
    \label{fig:apmCF}
    \vspace{-5pt}
\end{figure}

Affine progressive meshes allows us to losslessly recover the original meshes undergoing affine transformations. 
For smooth non-affine transformations such as our cube stylization, it could still be approximately recovered (see \reffig{apm}).
We summarize our cubic stylization with the affine progressive mesh in \refalg{fastCubeStyle}.
Note that the edge collapses is just a pre-processing step. 
In the online stage, one only needs to run cubic stylization on the coarsest mesh and then apply a sequence of vertex splits to visualize the result on the original resolution.
This offers a huge speed-up when interacting the parameter $λ$ on highly detailed models (see \reffig{apmCF}).

An interesting observation is that the number of faces $m$ in the coarsest mesh not only controls the runtime, but implicitly controls the frequency level of geometric details that gets preserved.
\begin{figure}
    \centering
    \includegraphics[width=3.33in]{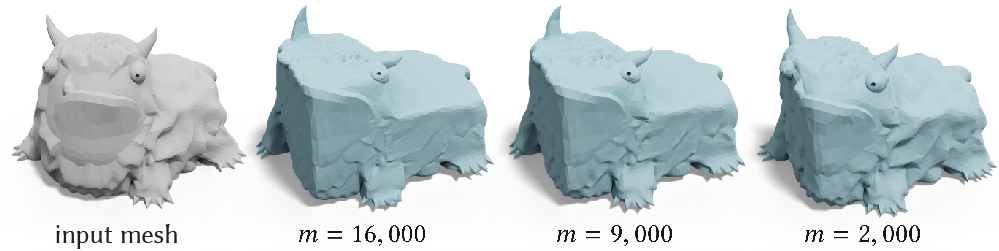}
    \caption{The number of faces $m$ used in the decimated mesh not only controls
    the runtime \update{but also} the frequency level of details that get preserved. \textcopyright Joseph Larson under CC BY.}
    \label{fig:changeFc}
    \vspace{-5pt}
\end{figure}  
In \reffig{changeFc} we show that, under the same $λ$, a smaller $m$ keeps details across a wider frequency range; in contrast, a larger $m$ only keeps details at higher frequencies.
Therefore one can manipulate the level of preserved features by playing with $m$.

\subsection{Implementation}
We implement the cubic stylization in C++ using \textsc{libigl} \cite{libigl} and evaluate our runtime on a MacBook Pro with an Intel i5 2.3GHz processor.
\reftab{runtime} lists the parameters and the runtime of our stylization in \reffig{rawCF} \update{(top)} and \reffig{apmCF}. 
We test our methods on meshes in the \textit{Thingi10K} \cite{Thingi10K} and show that we can obtain stylized geometry within a few seconds.
This is important for users to receive quick feedback on their parameter choices and iterate on their designs, such as the cubeness $λ$ in \reffig{changeLambda} and the the level of details $m$ in \reffig{changeFc}.
%

\paragraph{User study}
%
%
\begin{wrapfigure}[10]{r}{1.475in}
	\raggedleft
    \vspace{-9pt}
	\hspace*{-0.7\columnsep}
	\includegraphics[width=1.4in, trim={7mm 0mm 0mm 0mm}]{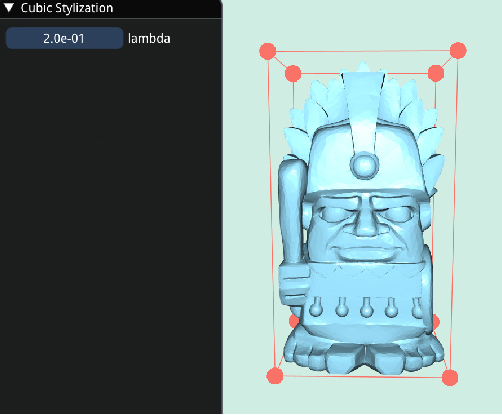} 
	\label{fig:UI}
\end{wrapfigure} 
We prototype a user interface (see the inset) to conduct an informal user study with six participants (4 male, 2 female) between the ages of 24 and 29. Participant 3D modeling experience ranged from none (complete novice) to three years of hobbyist use. Each participant was instructed for three minutes on how to use our software to load a mesh and control the cubeness parameter $λ$. Then we asked them to cubify a shape of their choosing from a collection of ten shapes. The results of their work is show in \reffig{userStudy}. All users reported that they were satisfied with the cubeness of their resulting shape. One user said that controlling the cubeness of their resulting shape is very easy because it only requires tuning a single parameter.
%
\begin{table}
    \centering
    \setlength{\tabcolsep}{5.425pt}
    \caption{
        For each example in \reffig{rawCF} and \reffig{apmCF}, we report the number of faces in the original model (\textsc{$|F|$}), $l1$ weight ($λ$), number of faces of the coarsest mesh ($m$), number of iterations (\textit{Iters.}), pre-processing time (\textit{Pre.}), and runtime at the online stage (\textit{Runtime}).
    }
    \begin{tabularx}{\linewidth}{lrrrrrr}
        \toprule
        \textit{Model} & \textsc{$|F|$} & $λ$ & $m$ &\textit{Iters.} & \textit{Pre.} & \textit{Runtime} \\
        \rowcolor{derekTableBlue}
        \reffig{rawCF}, left   & 39K & 0.20 & n/a & 106 & n/a &  $\mathbf{5.08s}$\\
        \reffig{rawCF}, mid. & 41K & 0.20 & n/a & 93 & n/a &  $\mathbf{4.50s}$\\
        \rowcolor{derekTableBlue}
        \reffig{rawCF}, right  & 21K & 0.4 & n/a & 86 & n/a &  $\mathbf{2.26s}$\\
        \reffig{apmCF}, left   & 2018K & 0.20 & 20K & 83 & 64.19$s$ &  $\mathbf{3.93s}$\\
        \rowcolor{derekTableBlue}
        \reffig{apmCF}, mid. & 346K & 0.40 & 20K & 222 & 10.69$s$ &  $\mathbf{4.59s}$\\
        \reffig{apmCF}, right  & 811K & 0.30 & 40K & 173 & 30.44$s$ &  $\mathbf{8.38s}$\\
        \bottomrule
    \end{tabularx}
    \smallskip
    \label{tab:runtime}
\end{table}
\begin{figure}
    \centering
    \includegraphics[width=3.33in]{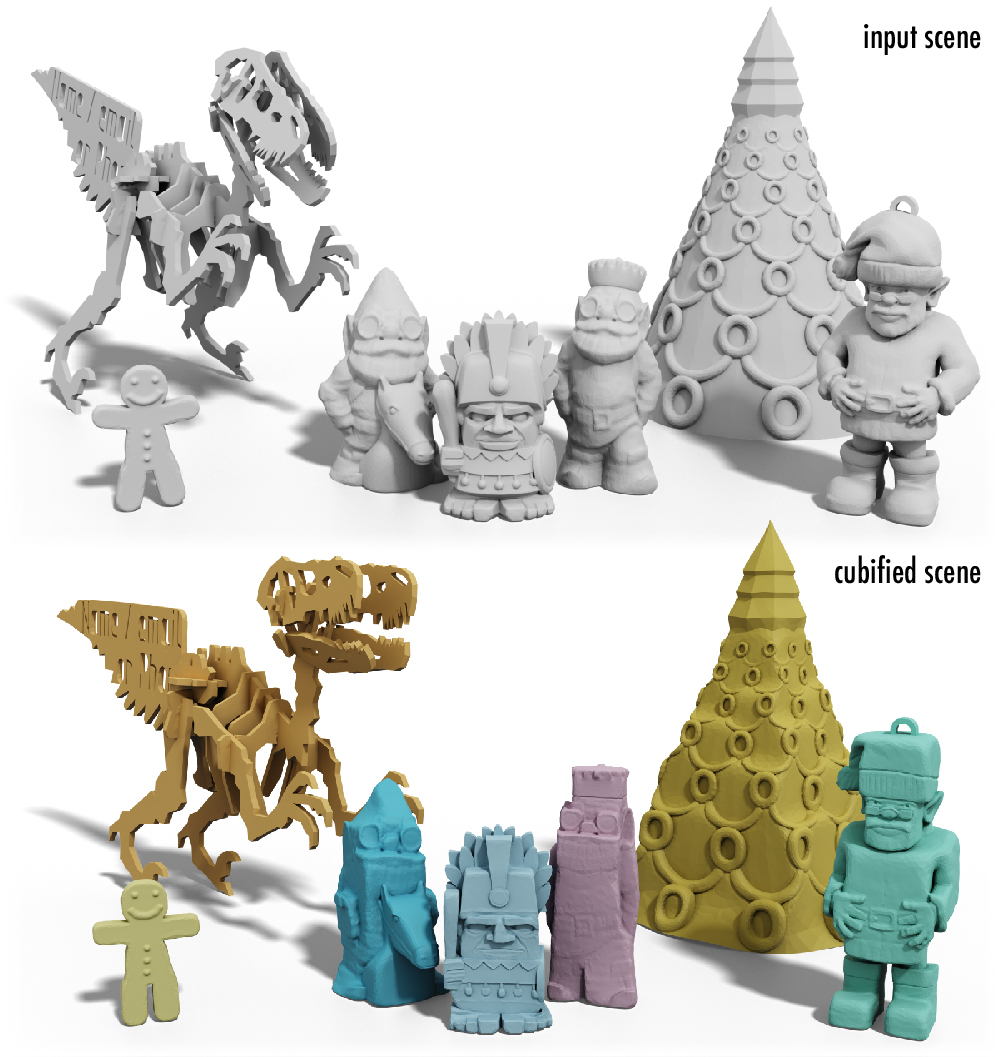}
    \caption{Even non-professional users can effortlessly turn an input scene (top) into a cubified scene (bottom). Different colors are results created by different users. From left to right, \textcopyright Peter Leppik, Cleven, TerenceKing, MakerBot, TerenceKing, PerryEngel, and Christina Chun under CC BY.}
    \label{fig:userStudy}
    \vspace{-5pt}
\end{figure}
\section{Artistic Controls}\label{sec:results}
%
\update{In addition to the two parameters $λ, m$, we expose many variants of our stylization to incorporate artistic controls. 
As a non-realistic modeling tool, this is important for users to realize their creativity. }

\begin{figure}
    \centering
    \includegraphics[width=3.33in]{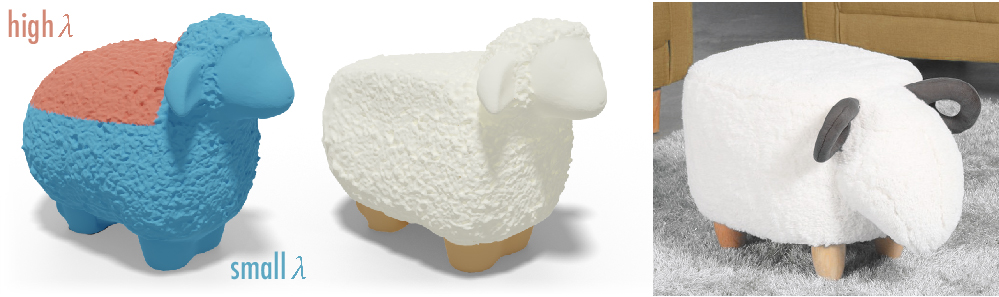}
    \caption{We vary $λ$ across the surface to have different cubeness for different parts. We apply higher $λ$ on the red region and smaller $λ$ for the blue region to create an ottoman-like shape (middle). \textcopyright pmoews under CC BY.}
    \label{fig:nonUniLambda}
    \vspace{-5pt}
\end{figure}

\begin{figure}
    \centering
    \includegraphics[width=3.33in]{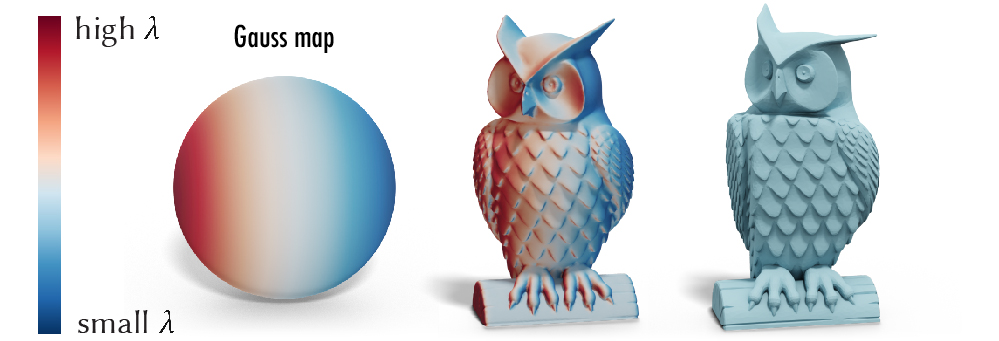}
    \caption{We can paint the $λ$ function on the Gauss map to have non-uniform $λ$ over the surface. In the figure, we have higher $λ$ for the original normals pointing towards left, and vice versa. \textcopyright Tom Cushwa under CC BY.}
    \label{fig:gauss}
    \vspace{-5pt}
\end{figure}

\begin{figure}
    \centering
    \includegraphics[width=3.33in]{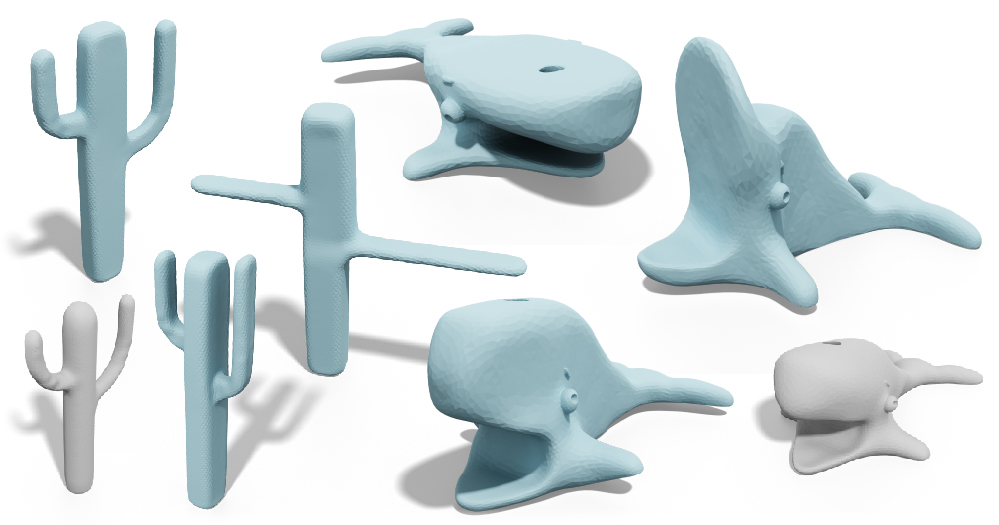}
    \caption{We can vary $λ$ for different axes to \update{turn inputs into the biased cubic style (blue) towards x,y,z axes} respectively. \textcopyright MakerBot (right) under CC BY.}
    \label{fig:biasAxis}
    \vspace{-5pt}
\end{figure}

We first focus our discussion on a variety of artistic controls that are related to the cubeness parameter $λ$. 
Although \refequ{cubeEnergy} only has a single $λ$ for an entire shape, we can actually specify different $λ_i$ for each vertex independently to have non-uniform cubeness, which leads to the expression $λ_i a_i \| \mR_i \vn_i \|_1$.
In \reffig{nonUniLambda}, we use this approach to make the back of the sheep much more cubic \update{than} the rest of the shape to create an ottoman-like geometry.
We can also specify the non-uniform cubeness $λ_i$ in a different way, instead of painting on the surface directly.
In \reffig{gauss} we \emph{paint} a function on the Gauss map in which the surface normal pointing towards left has higher cubeness.
When we map this function back to the surface, we can have a cubified owl that is more cubic when initial normals pointing towards the left and less cubic when pointing towards the right.
Similarly, we can have different $λ_x, λ_y, λ_z$ for different axes. 
In \reffig{biasAxis}, we replace the \cubeness in \refequ{cubeEnergy} with {$a_i (λ_x |(\mR_i \vn_i)_x| + λ_y |(\mR_i \vn_i)_y| + λ_z |(\mR_i \vn_i)_z|)$} and specify different values for each $λ_x, λ_y, λ_z$ to \update{have the style of a rectangular prism.}

\begin{figure}
    \centering
    \includegraphics[width=3.33in]{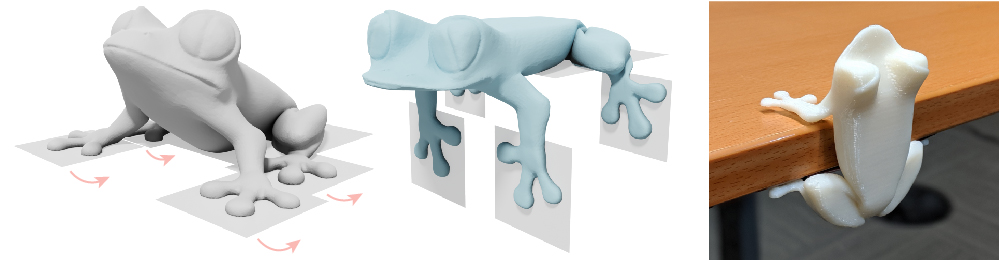}
    \caption{We constrain certain parts of the geometry lying on certain planes to create a 3D printed table clinger (right). \textcopyright Morena Protti under CC BY.}
    \label{fig:planeConst}
    \vspace{-5pt}
\end{figure}

If one wants to fix certain parts of the shape, we can easily add constraints in the global step, the same way as the method of \citet{sorkine2007rigid}. 
In \reffig{constraints} we add the parts constraint by fixing the position of some vertices when solving the linear system; we add the points constraint by specifying some deformed vertices $\U_i$ at user-desired positions.
We can also use the same methodology to constrain some parts of the geometry lying on certain planes.
For instance, setting $(\U_i)_x = 0$ can force vertex $i$ lying on the yz-plane.
In \reffig{planeConst} we use this plane constraint to create a table clinger.

\begin{figure}
    \centering
    \includegraphics[width=3.33in]{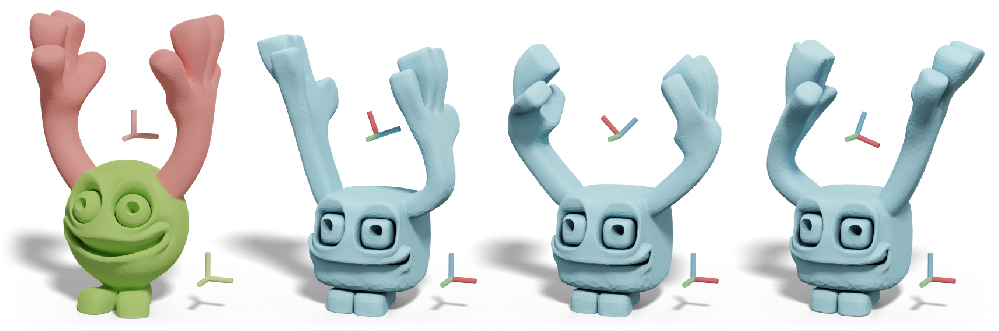}
    \caption{We can define the $\l1$-norm on different coordinate systems for different parts of the shape, instead of using the world coordinates. In the figure the hands and the body use different coordinate systems (left). By changing them, we can vary the cube orientations for different parts. \textcopyright David Hagemann under CC BY.}
    \label{fig:localFrames}
    \vspace{-5pt}
\end{figure}

\begin{figure}
    \centering
    \includegraphics[width=3.33in]{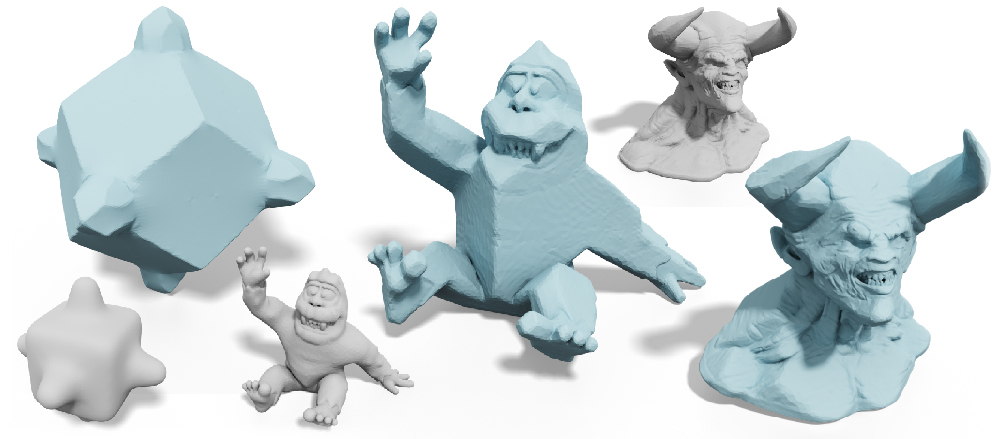}
    \caption{We apply a coordinate transformation inside the $\l1$-norm to generalize cubic stylization to polyhedrons. \textcopyright Proto Paradigm (middle), Ola Sundberg (right) under CC BY.}
    \label{fig:polyCF}
    \vspace{-5pt}
\end{figure}

\begin{figure}
    \centering
    \includegraphics[width=3.33in]{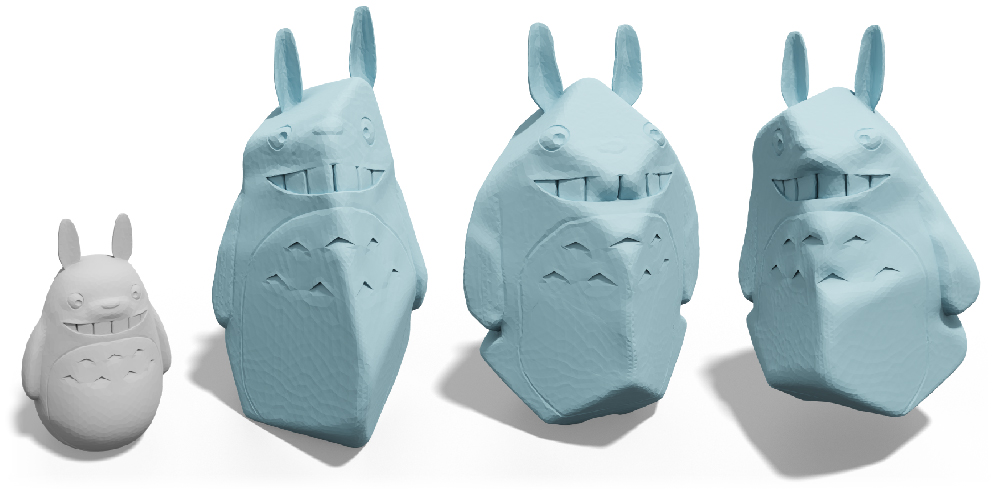}
    \caption{We apply non-symmetric coordinate transformations inside the $\l1$-norm to create irregular polyhedral stylization. \textcopyright Johannes under CC BY.}
    \label{fig:irregularPoly}
    \vspace{-5pt}
\end{figure}

\begin{figure}
    \centering
    \includegraphics[width=3.33in]{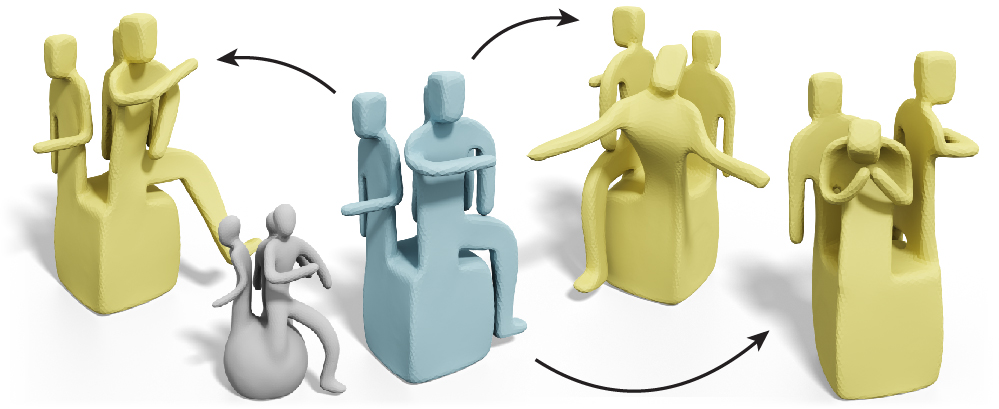}
    \caption{Once we have the cubic geometry (blue), standard deformation techniques (e.g., \cite{sorkine2007rigid}) can be used to manipulate the cubified shape (yellow).}
    \label{fig:postDeform}
    \vspace{-5pt}
\end{figure}

In addition, one can utilize the property of the $\l1$-norm to have different artistic effects. 
Because the \cubeness term is orientation dependent, in \reffig{orientations} we can apply different rotations to the mesh before the stylization to control the results. 
Rather than rotating the mesh, another way is to encode the normal vector in a different coordinate system $λ a_i \| \mR_i \vn_i^\text{local} \|_1$, where we use $\vn_i^\text{local}$ to denote the user-desired coordinate system for vertex $i$.
This perspective allows us to define the $\l1$-norm on different coordinate systems for different parts of the shape to obtain different cube orientations (\reffig{localFrames}).
Beyond the cubic stylization, in \update{\reffig{polyCF}, \reffignum{irregularPoly}} we apply a coordinate transformation $\mB$ inside the $\l1$-norm $λ a_i \| \mB \mR_i \vn_i \|_1$ to achieve polyhedral stylization, for which we provide the details in \refapp{polyGeneralization}.
Once we obtain the stylized shapes, they are ready to be used by standard deformation techniques in animations (\reffig{postDeform}).

\section{Limitations \& Future Work}\label{sec:limitations}
\begin{figure}
    \centering
    \includegraphics[width=3.33in]{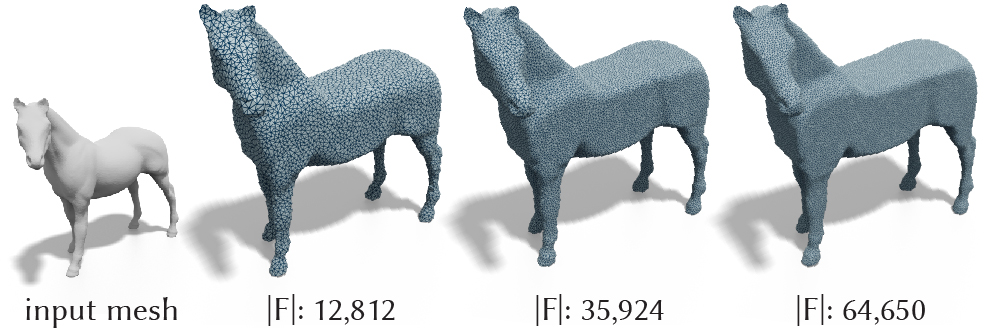}
    \caption{\update{Although exhibiting similar cubenesses, our stylization is still not invariant to different resolutions.}}
    \label{fig:diffRes}
    \vspace{-5pt}
\end{figure}

Accelerating the stylization to real-time would enable faster iterations between designs.
Developing a more robust stylization to for bad quality triangles, non-manifold meshes, or even point cloud could be useful for stylizing real-world geometric data.
\update{Guaranteeing results to be self-intersection free would be desirable for downstream tasks.}
\update{Extending our energy to be invariant to discretizations could achieve more consistent results across different resolutions (see \reffig{diffRes}).}
Extending to quadrilateral meshes and NURBS surfaces could benefit existing modeling or engineering design softwares. 
\update{Generalizing to volumetric meshes could have a better volume preservation.}
Exploring different deformation energies and $\ell^p$-norm could lead to novel stylization tools for non-realistic modeling. 
Beyond generating stylized shapes, the mathematical expression of the cubic geometry could offer insights toward understanding more intricate styles.
For instance, \textit{Cubism} has been considered as a revolutionized artistic style for paintings and sculptures.
Cubism has appeared since the early 20th century.
%
Since then, several attempts have tried to describe \cite{henderson1983fourth} and generate Cubist art \cite{wang2011cubist, corker20184d}, 
%
but more efforts still required to offer scientific explanations to a wide variety of Cubist art.
Our cubic stylization only focuses on a specific style. 
We hope this could inspire future attempts to capture different sculpting styles such as those \update{presented} in African art, or even a generic approach to create different styles in an unified framework.

\begin{acks}
%
Our research is funded in part by New Frontiers of Research Fund (NFRFE–201), the Ontario Early Research Award program, NSERC Discovery (RGPIN2017–05235, RGPAS–2017–507938), the Canada Research Chairs Program, the Fields Centre for Quantitative Analysis and Modelling and gifts by Adobe Systems, Autodesk and MESH Inc. We thank members of Dynamic Graphics Project at the University of Toronto; Michael Tao and Wen-Hsiang Tsai for project motivations; David I.W. Levin and Yotam Gingold for ideas on the artistic controls and the user study; Oded Stein for sharing results; Rahul Arora for fabricating stylized shapes; Leonardo Sacht and Silvia Sell\'an for proofreading; Omid Poursaeed, Rahul Arora, Whitney Chiu, Yang Zhou, Yifan Wang, Youssef Alami Mejjati, and Zhicong Lu for participating in the user study.
\end{acks}

\bibliographystyle{ACM-Reference-Format}
\bibliography{sections/reference}
\appendix

\section{Polyhedral Generalization}\label{app:polyGeneralization}
\begin{figure}[h]
    \centering
    \includegraphics[width=3.33in]{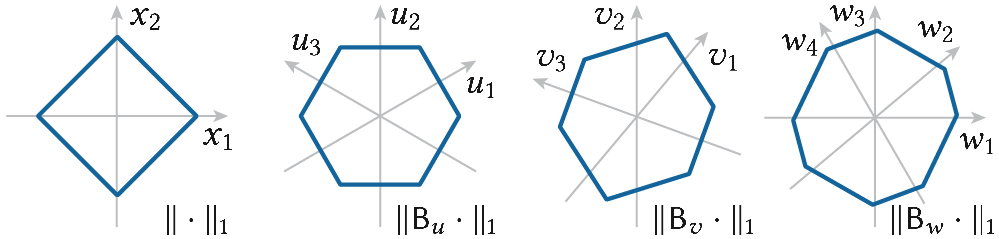}
    \caption{\update{By specifying different coordinate transformations $\mB$ inside the $\l1$-norm, we can encourage polyhedral style.}}
    \label{fig:2Dpoly}
    \vspace{-5pt}
\end{figure}
Simply applying a coordinate transformation $B:\R^n → \R^m$ inside the $\l1$-norm can encourage polyhedral results, instead of cubic results (see \reffig{2Dpoly}). \update{The $\l1$-norm of a vector is defined as the summation of its magnitudes along each basis vector. Thus applying a coordinate transformation inside the $\l1$-norm changes its bahavior because the basis vectors are different.} Following the notation in \refequ{cubeEnergy}, polyhedron energy can be written as 
\begin{align*}
    \minimize_{\U, \{ \mR_i\} }\ \sum_{i∈V} \sum_{j∈ \mathcal{N}(i)} \frac{w_{ij}}{2}  \| \mR_i \dV_{ij} - \dU_{ij} \|^2_F + λ a_i \| \mB \mR_i \vn_i \|_1.
\end{align*}
\update{In our case,} $\mB$ is a $m$-by-$3$ coordinate transformation matrix for shapes embedded in $\R^3$. Again by setting $\vz = \mR_i \vn_i$ we can reach almost the same optimization procedures, except \update{the} \refequ{ZStep} now becomes (we ignore the iteration superscript for clarity)
\begin{align}
	\vz^{k+1} ← \argmin_{\vz}\ λ a_i \| \mB \vz \|_1 +  \frac{ρ}{2} \| \mR_i\vn_i - \vz + \vu \|^2_2. \label{equ:polyZStep}
\end{align}
\update{Similar to common techniques for solving the \emph{Basis Pursuit} problem,} we introduce a variable $\vt \succeq \| \mB \vz \|_1$ to transform \refequ{polyZStep} into a small quadratic program subject to equality constraints
\begin{alignat*}{2}
	&\minimize_{\vz, \vt}\ 
	&&\begin{bmatrix}
        \vz^⊤\ \vt^⊤ 
	\end{bmatrix}
	\begin{bmatrix}
		\nicefrac{ρ}{2}\cdot \matFont{I}_3 &\matFont{0} \\
		\matFont{0} &\matFont{0}
	\end{bmatrix}
	\begin{bmatrix}
        \vz \\ \vt
	\end{bmatrix} \\
	& &&\qquad
	+ 
	\begin{bmatrix}
        -ρ(\mR_i\vn_i+ \vu)^⊤ \ λ a_i \matFont{1}_m^⊤ 
	\end{bmatrix}
	\begin{bmatrix}
        \vz \\ \vt
	\end{bmatrix}\\
	&\text{ subject to}\ 
	&&\begin{bmatrix}
		\mB &-\matFont{I}_m \\
		-\mB &-\matFont{I}_m
	\end{bmatrix}
	\begin{bmatrix}
        \vz \\ \vt
	\end{bmatrix}
	\preceq \matFont{0},
	\label{equ:QP}
\end{alignat*}
\update{where $\matFont{I}_x$ and $\matFont{1}_x$ denote the identity matrix with size $x$ and a column vector of $1$ with size $x$ respectively.} We then solve this efficiently using \textsc{cvxgen} \cite{mattingley2012cvxgen}. Note that the results in \reffig{polyCF} and \reffig{irregularPoly} use $m=4$.

\end{document}